\documentstyle[twoside,epsfig]{article}

\catcode`\@=11
\long\def\@makefntext#1{
\protect\noindent \hbox to 3.2pt {\hskip-.9pt
$^{{\eightrm\@thefnmark}}$\hfil}#1\hfill}		

\def\@makefnmark{\hbox to 0pt{$^{\@thefnmark}$\hss}}	

\def\ps@myheadings{\let\@mkboth\@gobbletwo
\def\@oddhead{\hbox{}
\rightmark\hfil\eightrm\thepage}
\def\@oddfoot{}\def\@evenhead{\eightrm\thepage\hfil
\leftmark\hbox{}}\def\@evenfoot{}
\def\sectionmark##1{}\def\subsectionmark##1{}}

\oddsidemargin=\evensidemargin
\newcounter{sectionc}\newcounter{subsectionc}\newcounter{subsubsectionc}
\renewcommand{\section}[1] {\vspace{12pt}\addtocounter{sectionc}{1}
\setcounter{subsectionc}{0}\setcounter{subsubsectionc}{0}\noindent
	{\tenbf\thesectionc. #1}\par\vspace{5pt}}
\renewcommand{\subsection}[1] {\vspace{12pt}\addtocounter{subsectionc}{1}
	\setcounter{subsubsectionc}{0}\noindent
	{\bf\thesectionc.\thesubsectionc. {\kern1pt \bfit #1}}\par\vspace{5pt}}
\renewcommand{\subsubsection}[1] {\vspace{12pt}\addtocounter{subsubsectionc}{1}
	\noindent{\tenrm\thesectionc.\thesubsectionc.\thesubsubsectionc.
	{\kern1pt \tenit #1}}\par\vspace{5pt}}
\newcommand{\nonumsection}[1] {\vspace{12pt}\noindent{\tenbf #1}
	\par\vspace{5pt}}

\newcounter{appendixc}
\newcounter{subappendixc}[appendixc]
\newcounter{subsubappendixc}[subappendixc]
\renewcommand{\thesubappendixc}{\Alph{appendixc}.\arabic{subappendixc}}
\renewcommand{\thesubsubappendixc}
	{\Alph{appendixc}.\arabic{subappendixc}.\arabic{subsubappendixc}}

\renewcommand{\appendix}[1] {\vspace{12pt}
	  \refstepcounter{appendixc}
	  \setcounter{figure}{0}
	  \setcounter{table}{0}
	  \setcounter{lemma}{0}
	  \setcounter{theorem}{0}
	  \setcounter{corollary}{0}
	  \setcounter{definition}{0}
	  \setcounter{equation}{0}
	  \renewcommand{\thefigure}{\Alph{appendixc}.\arabic{figure}}
	  \renewcommand{\thetable}{\Alph{appendixc}.\arabic{table}}
	  \renewcommand{\theappendixc}{\Alph{appendixc}}
	  \renewcommand{\thelemma}{\Alph{appendixc}.\arabic{lemma}}
	  \renewcommand{\thetheorem}{\Alph{appendixc}.\arabic{theorem}}
	  \renewcommand{\thedefinition}{\Alph{appendixc}.\arabic{definition}}
	  \renewcommand{\thecorollary}{\Alph{appendixc}.\arabic{corollary}}
	  \renewcommand{\theequation}{\Alph{appendixc}.\arabic{equation}}
	  \noindent{\tenbf Appendix \theappendixc #1}\par\vspace{5pt}}
\newcommand{\subappendix}[1] {\vspace{12pt}
	  \refstepcounter{subappendixc}
	  \noindent{\bf Appendix \thesubappendixc. {\kern1pt \bfit #1}}
	\par\vspace{5pt}}
\newcommand{\subsubappendix}[1] {\vspace{12pt}
	  \refstepcounter{subsubappendixc}
	  \noindent{\rm Appendix \thesubsubappendixc. {\kern1pt \tenit #1}}
	\par\vspace{5pt}}

\newcommand{\textlineskip}{\baselineskip=13pt}
\newcommand{\smalllineskip}{\baselineskip=10pt}

\def\eightcirc{
\begin{picture}(0,0)
\put(4.4,1.8){\circle{6.5}}
\end{picture}}
\def\eightcopyright{\eightcirc\kern2.7pt\hbox{\eightrm c}}

\newcommand{\copyrightheading}[1]
      {\vspace*{-2.5cm}\smalllineskip{\flushleft
      {\footnotesize International Journal of Modern Physics B, #1}\\
      {\footnotesize $\eightcopyright$\, World Scientific Publishing
       Company}\\
       }}

\newcommand{\publisher}[2]{{\begin{center}\footnotesize\smalllineskip
      Received #1\\
      Revised #2
      \end{center}
      }}

\def\abstracts#1#2#3{{
      \centering{\begin{minipage}{4.5in}\baselineskip=10pt\footnotesize
      \parindent=0pt #1\par
      \parindent=15pt #2\par
      \parindent=15pt #3
      \end{minipage}}\par}}

\renewenvironment{thebibliography}[1]                 
      {\frenchspacing
       \ninerm\baselineskip=11pt
       \begin{list}{\arabic{enumi}.}
      {\usecounter{enumi}\setlength{\parsep}{0pt}
       \setlength{\leftmargin 12.7pt}{\rightmargin 0pt} 
       \setlength{\itemsep}{0pt} \settowidth
      {\labelwidth}{#1.}\sloppy}}{\end{list}}

\newcounter{itemlistc}
\newcounter{romanlistc}
\newcounter{alphlistc}
\newcounter{arabiclistc}

\newcommand{\fcaption}[1]{
        \refstepcounter{figure}
        \setbox\@tempboxa = \hbox{\footnotesize Fig.~\thefigure. #1}
        \ifdim \wd\@tempboxa > 5in
           {\begin{center}
        \parbox{5in}{\footnotesize\smalllineskip Fig.~\thefigure. #1}
            \end{center}}
        \else
             {\begin{center}
             {\footnotesize Fig.~\thefigure. #1}
              \end{center}}
        \fi}

\newcommand{\tcaption}[1]{
        \refstepcounter{table}
        \setbox\@tempboxa = \hbox{\footnotesize Table~\thetable. #1}
        \ifdim \wd\@tempboxa > 5in
           {\begin{center}
        \parbox{5in}{\footnotesize\smalllineskip Table~\thetable. #1}
            \end{center}}
        \else
             {\begin{center}
             {\footnotesize Table~\thetable. #1}
              \end{center}}
        \fi}

\def\@citex[#1]#2{\if@filesw\immediate\write\@auxout
      {\string\citation{#2}}\fi
\def\@citea{}\@cite{\@for\@citeb:=#2\do
      {\@citea\def\@citea{,}\@ifundefined
      {b@\@citeb}{{\bf ?}\@warning
      {Citation `\@citeb' on page \thepage \space undefined}}
	{\csname b@\@citeb\endcsname}}}{#1}}

\newif\if@cghi
\def\cite{\@cghitrue\@ifnextchar [{\@tempswatrue
	\@citex}{\@tempswafalse\@citex[]}}
\def\citelow{\@cghifalse\@ifnextchar [{\@tempswatrue
	\@citex}{\@tempswafalse\@citex[]}}
\def\@cite#1#2{{$\null^{#1}$\if@tempswa\typeout
	{IJCGA warning: optional citation argument
	ignored: `#2'} \fi}}

\def\pmb#1{\setbox0=\hbox{#1}
	\kern-.025em\copy0\kern-\wd0
	\kern.05em\copy0\kern-\wd0
	\kern-.025em\raise.0433em\box0}

\def\fnt#1#2{\footnotetext{\kern-.3em
	{$^{\mbox{\scriptsize #1}}$}{#2}}}

\def\fpage#1{\begingroup
\voffset=.3in
\thispagestyle{empty}\begin{table}[b]\centerline{\footnotesize #1}
	\end{table}\endgroup}

\def\runninghead#1#2{\pagestyle{myheadings}
\markboth{{\protect\footnotesize\it{\quad #1}}\hfill}
{\hfill{\protect\footnotesize\it{#2\quad}}}}
\font\tenrm=cmr10
\font\tenit=cmti10
\font\tenbf=cmbx10
\font\bfit=cmbxti10 at 10pt
\font\ninerm=cmr9
\font\nineit=cmti9
\font\ninebf=cmbx9
\font\eightrm=cmr8

\def\qed{\hbox{${\vcenter{\vbox{			
   \hrule height 0.4pt\hbox{\vrule width 0.4pt height 6pt
   \kern5pt\vrule width 0.4pt}\hrule height 0.4pt}}}$}}

\def\bsc{{\sc a\kern-6.4pt\sc a\kern-6.4pt\sc a}}	
\def\bflatex{\bf L\kern-.30em\raise.3ex\hbox{\bsc}\kern-.14em
T\kern-.1667em\lower.7ex\hbox{E}\kern-.125em X}

\begin{document}

\runninghead{Pseudo-gap features of intrinsic tunneling in
(HgBr$_2$)-Bi2212 single crystals} {Pseudo-gap features of intrinsic tunneling in
(HgBr$_2$)-Bi2212 single crystals}

\normalsize\textlineskip
\thispagestyle{empty}
\setcounter{page}{1}

\copyrightheading{}	

\vspace*{0.88truein}

\fpage{1}
\centerline{\bf PSEUDO-GAP FEATURES OF INTRINSIC TUNNELING IN}
\vspace*{0.035truein}
\centerline{\bf (HgBr$_2$)-Bi2212 SINGLE CRYSTALS
\footnote{Supported by the Swedish Superconductivity Consortium and
the Royal Swedish Academy of Sciences}}
\vspace*{0.37truein}
\centerline{\footnotesize AUGUST YURGENS,\footnote{On leave from
P.L. Kapitza Institute, Moscow, Russia; e-mail: yurgens@fy.chalmers.se}
\hspace*{2pt} DAG WINKLER, and TORD CLAESON}
\vspace*{0.015truein}
\centerline{\footnotesize\it Department of Microelectronics 
and Nanoscience, Chalmers University of Technology}
\baselineskip=10pt
\centerline{\footnotesize\it and G\"{o}teborg University, S-41296, 
G\"{o}teborg, Sweden}
\vspace*{10pt}
\centerline{\footnotesize SEONG-JU HWANG and JIN-HO CHOY}
\vspace*{0.015truein}
\centerline{\footnotesize\it Department of Chemistry,
Seoul National University,}
\baselineskip=10pt
\centerline{\footnotesize\it	Seoul 151-742, Korea}
\vspace*{0.225truein}
\publisher{(received date)}{(revised date)}

\vspace*{0.21truein}
\abstracts{The c-axis tunneling properties of both pristine Bi2212 and
its HgBr$_2$ intercalate have been measured in the temperature
range 4.2 - 250~K. Lithographically patterned 7-10 unit-cell
heigh mesa structures on the surfaces of these single crystals
were investigated. Clear SIS-like tunneling curves for current
applied in the $\it c$-axis direction have been observed. The dynamic
conductance d$I/$d$V(V)$
shows both sharp peaks corresponding to a
superconducting gap edge and a dip feature beyond the gap, followed
by a wide maximum, which persists up to a room temperature.
Shape of the temperature dependence of the {\it c}-axis resistance
does not change after the intercalation suggesting
that a coupling between $\rm CuO_2$-bilayers
has little effect on the pseudogap. 
}{}{}



\vspace*{1pt}\textlineskip	 
\section{Introduction}	 
\vspace*{-0.5pt}
\noindent

The existence of a pseudogap in electronic excitation spectra
is believed to be one of the very important features of 
high-$T_c$ (HTS) superconductors.\cite{all} 
This gap was
reported to exist in underdoped samples, although some
experiments indicate that the pseudogap is present in
overdoped samples also.\cite{overdoped,STM}

The tunneling spectroscopy
is particular sensitive to the density
of states (DOS) at the Fermi level and therefore
can be used to study any gap in the quasiparticle excitation
spectrum.  The most common experimental methods, STM and
point-contact techniques are however essentially surface probes.
The surfaces of most HTS compounds	
deteriorates with time unless special measures are undertaken.
That is why there is a great variety of data quality
reported in the literature.\cite{Hasegawa}

It is now experimentally established that highly anisotropic
layered HTS exhibit intrinsic tunneling effects.\cite{Muller}
For example, in Bi2212 ($\rm Bi_2Sr_2CaCu_2O_8$) the
$\approx 3$~\AA \ thick metallic copper oxide
sheets (CuO-bilayers] are separated by $\approx $12 \AA \ thick 
insulating layers.
Out-of-plane (or {\it c}-axis) charge transport 
occurs via a sequential 
tunneling
of electrons or Cooper pairs between these sheets.\cite{Muller}
The atomic perfection of the naturally occurring tunnel junctions
in such materials provides a reliable basis for the tunneling
spectroscopy \textit{inside} the single crystal giving
a high degree of homogeneity and reproducibility.

The phonon-mediated sub-gap structures of such intrinsic Josephson 
junctions
(IJJ) has already been  studied.\cite{sub_gap}
The pseudogap spectroscopy
with intrinsic tunnel junctions has also been 
attempted.\cite{Suzuki,Gough}

\begin{figure}[!h]
\vspace{-3.5cm}
\hspace{5cm}
\epsfig{file=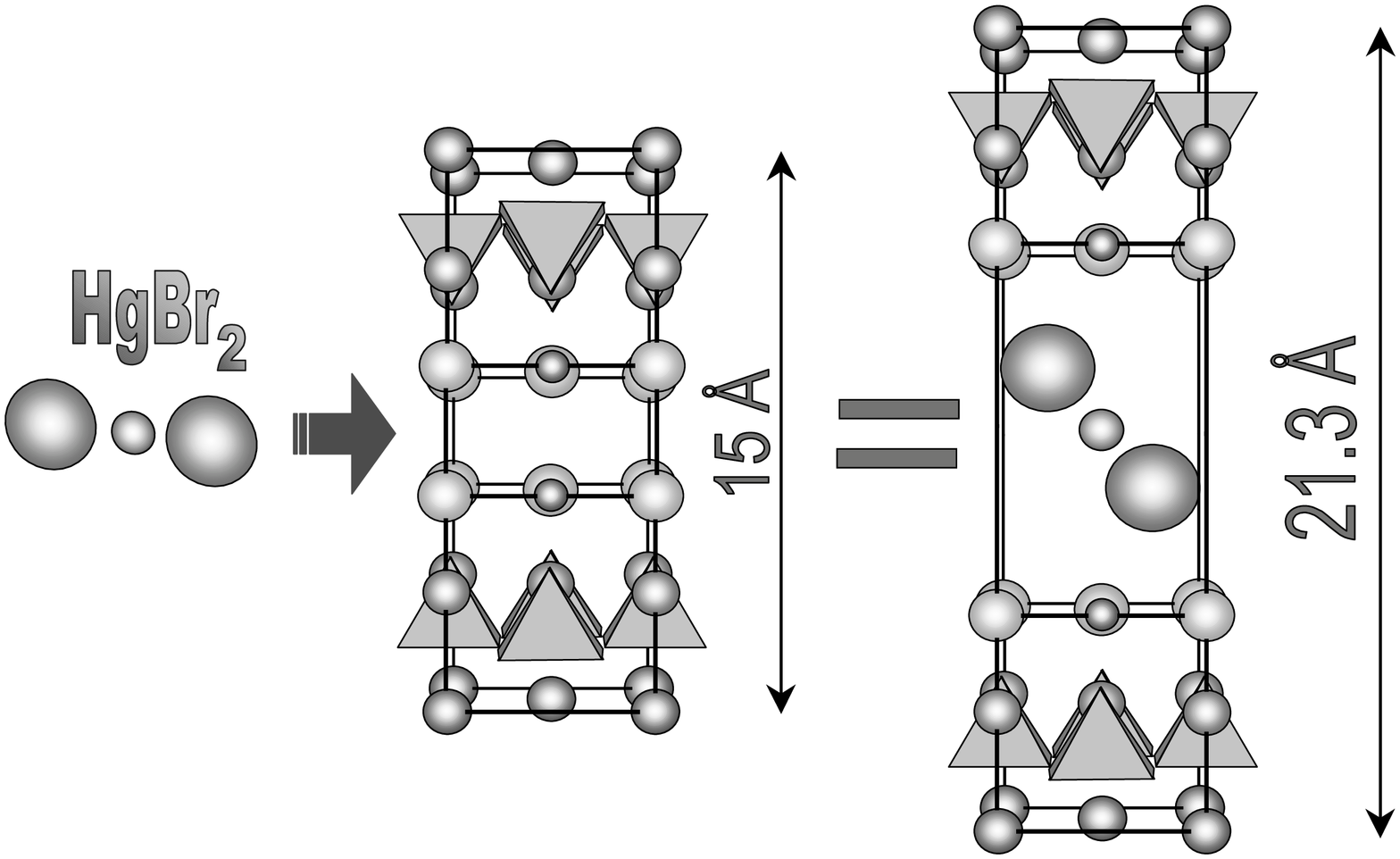,height=6.0cm,angle=90}
\fcaption{Schematic picture of $\rm HgBr_2$-intercalation.}
\label{structure}
\end{figure}

Here, we repeat experiments of Refs.\cite{Suzuki,Gough} on
HgBr$_2$-Bi2212 samples. Insertion of inert 
HgBr$_2$-molecules in between adjacent BiO-layers 
results in a significant stretching of Bi2212 crystals in the
\textit{c}-axis direction, see Fig.~\ref{structure} without affecting
the superconducting critical temperature $T_c$ much. 
The intercalation leads to decrease in
the \textit{c}-axis critical current $I_c$ and increase in the normal
state resistance $R_c$.	 The Joule heating,
which was considered to be a problem in
experiments on IJJ at high bias current,\cite{Suzuki,Gough}
can therefore be significantly suppressed.
Moreover, the effect of the interlayer coupling on the 
pseudogap can be investigated.

\begin{figure}[!h]
\begin{center}
\epsfig{file=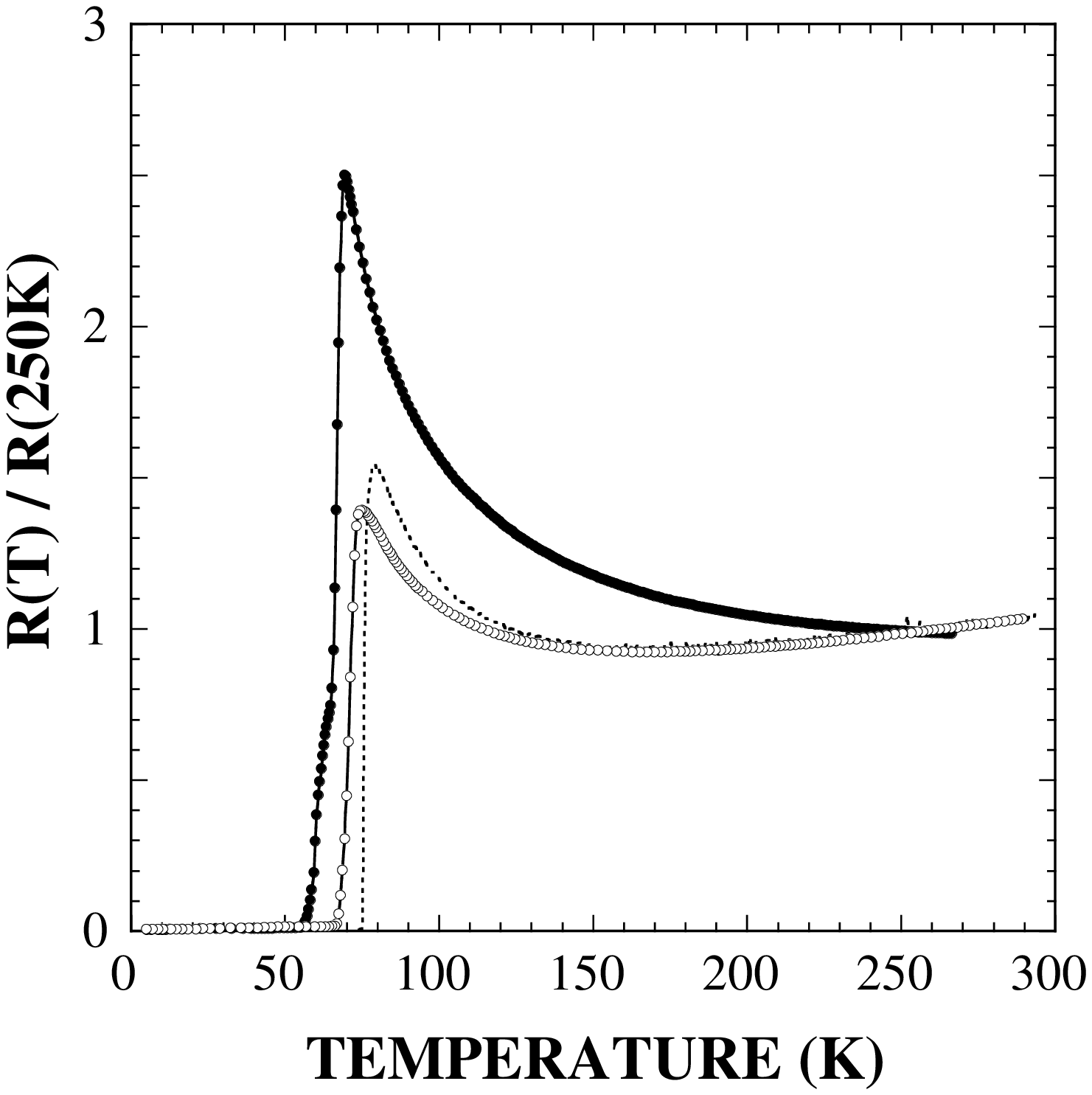,height=6.0cm}
\end{center}
\vspace{5mm}
\fcaption{The temperature dependence of the {\it c}-axis resistance of
the pristine (dash line) and intercalated samples (circles and lines).}
\label{RT}
\end{figure}

\section{Experiment}
\noindent
Bi2212 single crystals were synthesized by the traveling-zone method.
The intercalation of $\rm HgBr_2$ was perfomed by heating the vacuum 
sealed tube containing pristine single crystals and 
$\rm HgBr_2$ during about one week.\cite{Choy} According to
X-ray analysis, the lattice expansion after the intercalation
was $\sim 12.6$~\AA.

\begin{figure}[!h]
\begin{center}
\epsfig{file=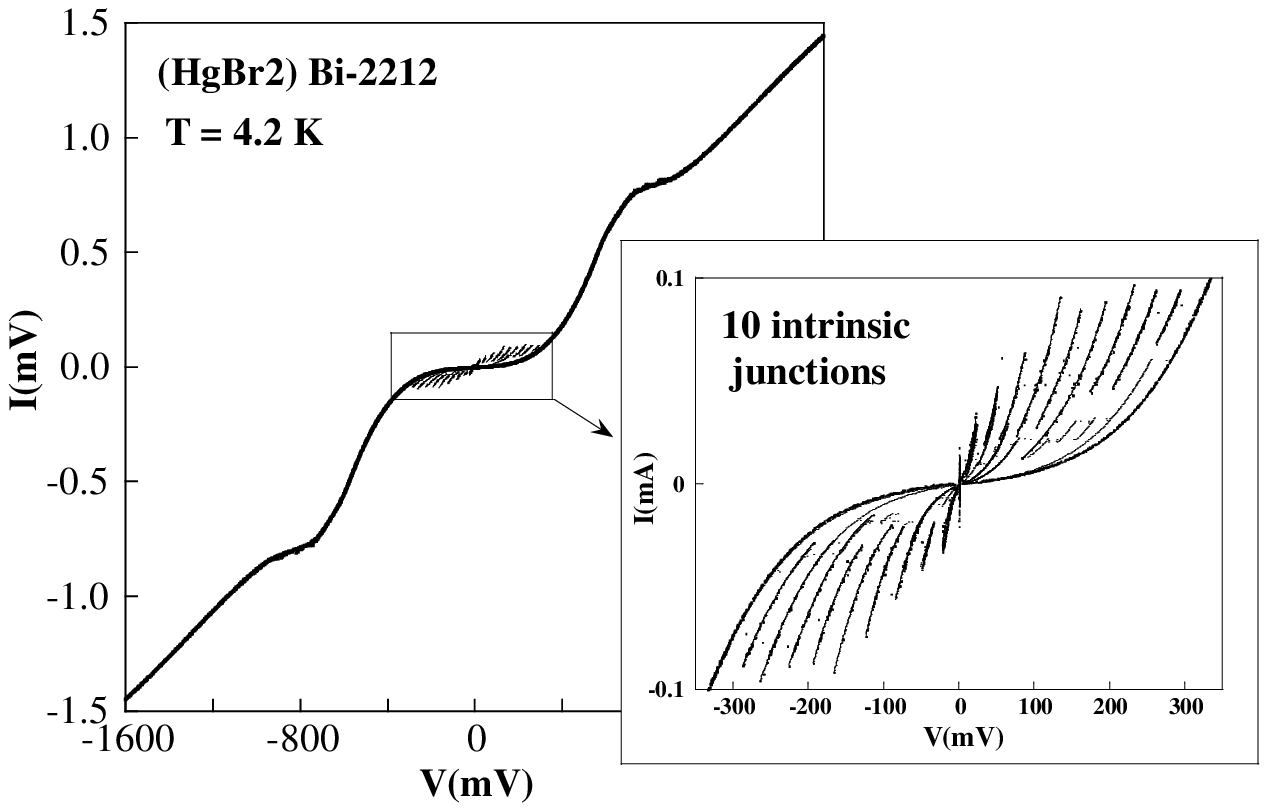,height=6.0cm}
\end{center}
\vspace{5mm}
\fcaption{Current-voltage characteristics for
$\rm (HgBr_2$)Bi2212- sample. The contact resistance of 5~$\Omega $
has not been subtracted.}
\label{fig1}
\end{figure}

To investigate the \textit{c}-axis transport properties, several
mesas with areas $\rm 10-600\ \mu m^2$ and heights 
$150-200$~\AA \ %
were fabricated on the freshly-cleaved surfaces of 
the single crystals 
using a standard photolithography and Ar-ion milling.\cite{we_APL}
The resulting mesas comprised $\sim 5-15$ IJJ in series.

\section{Results and discussion}
\noindent
Fig.~\ref{RT} shows the temperature dependence of 
 $R_c$ for both the pristine (a dash line) and 
intercalated samples (symbols and lines). 
The intercalation lead to a significant
increase of the {\it c}-axis resistivity $\rho $.  
At $T\approx 175$~K it is
about $\sim $10~$\Omega $~cm for the former, 
and  $\sim $2~k$\Omega $~cm for the latter samples. 
The pristine samples were slightly overdoped after annealing
in oxygen, which is seen
both from the reduced value of $T_c\approx 75$~K, and from the 
temperature dependence of the resistance showing the
linearly increasing with temperature part at high
temperatures, see Fig.~\ref{RT}. It
can be also seen that the linear part of $R_c(T)$ disappeared
after the intercalation in one sample, while conserved in another.  
After intercalation, $T_c$ somewhat further decreased,
down to about 62-65~K. $T_c$ of Bi2212 seems to be less
prone to HgBr$_2$-molecules as compared with other
intercalates, despite larger stretching of the 
lattice.\cite{Choy,iodine} 

Recently, it has been shown that the onset of the 
upturn in $R_c(T)$  correlates with 
appearance of the non-linearities in the {\it c}-axis
tunneling characteristics with decreasing temperature.\cite{Suzuki}
These non-linearities bear witness of pseudogap features
in the {\it c}-axis transport. 
From our experiments we 
see that the intercalation does not influence
the characteristic temperature $T^*$ in
a straightforward way. Despite significant
increase, the {\it c}-axis resistivity can preserve the form of the
temperature dependence. This implies that the out-of-plane
resistance and pseudogap features are determined mainly by
properties of individual CuO-bilayers in Bi-2212. The coupling between 
different bilayers
seems has a little effect on pseudogap, 
as long as oxygen content is preserved.\cite{we_PRL2}

\begin{figure}[!h]
\begin{center}
\epsfig{file=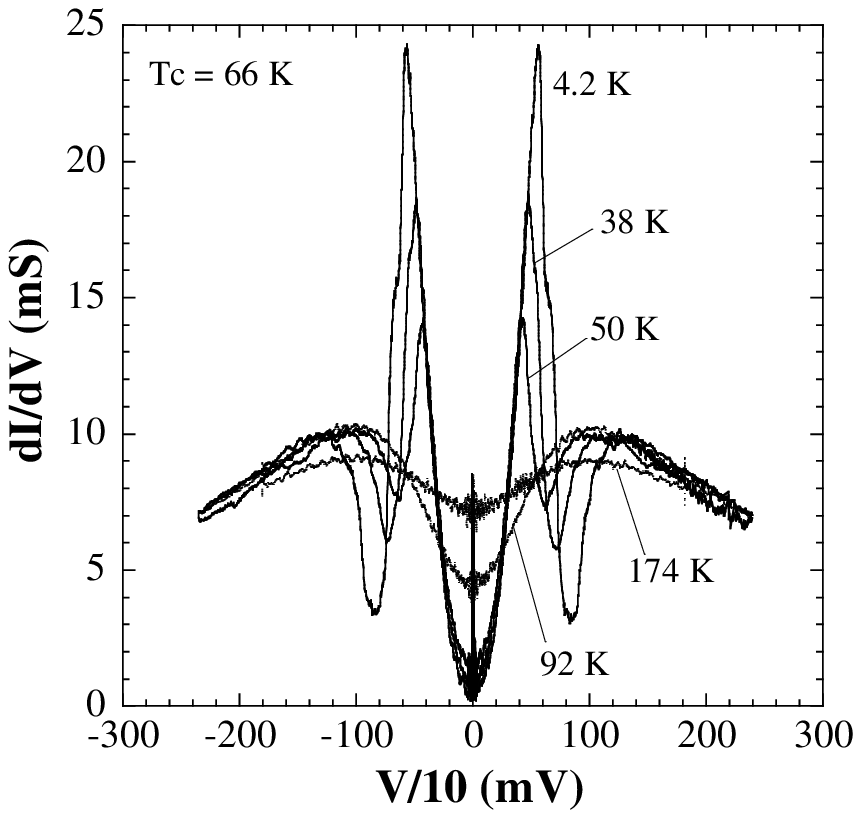,width=7.0cm}
\end{center}
\fcaption{d$I$/d$V(V)$ at different temperatures for
$\rm (HgBr_2$)Bi2212- sample.}
\label{fig2}
\end{figure}

The main frame of Fig.~\ref{fig1} shows a typical $I-V$-characteristic
of a mesa with 10 IJJ. The inset shows an enlarged central part of the 
$I-V$-characteristic.  
The majority of junctions have $I_c\sim 0.1$~mA, which
is more than an order of magnitude smaller than $I_c$ of the
 pristine-crystal  mesas with
the same area $A$ ($\rm \sim 200\ \mu m$).\cite{we_APL}  
The normal state
tunneling parts of the $I-V$-characteristics are clearly seen at 
$\sim $1~mA.
The dynamic conductance d$I/$d$V$ vs. the average voltage per one IJJ 
in the mesa was deduced by numerical differentiation 
of the $I-V$ curve corresponding
to all 10 IJJ being in the quasiparticle state, Fig.~\ref{fig2}. 
d$I/$d$V$ of 0.01~S
at $\sim $100~mV in our 
experiments would correspond to $10^{-10}$~S for typical STM-contact
areas (say, 15$\times $15~\AA$^2$), emphasizing a high quality of the 
tunneling barriers in IJJ.

Sharp peaks seen in Fig.~\ref{fig2} likely correspond 
to the superconducting
energy gap $V_g$ of $\rm CuO_2$ layers.  
The dips at $\sim 1.5V_g$ were
attributed to strong-coupling effects.\cite{deWilde} 
A gently sloping dip at the zero voltage at $T>T_c$ likely represents 
the pseudogap in the DOS.  It has the almost temperature independent 
energy scale of $\sim 100$~mV per each IJJ, which is somewhat larger 
than in experiments of Suzuki \textit{et al.}\cite{Suzuki} 
It is gradually smearing with temperature, although being 
seen in $I-V$'s at temperatures up to about 250~K for some samples.

\begin{figure}[!h]
\begin{center}
\epsfig{file=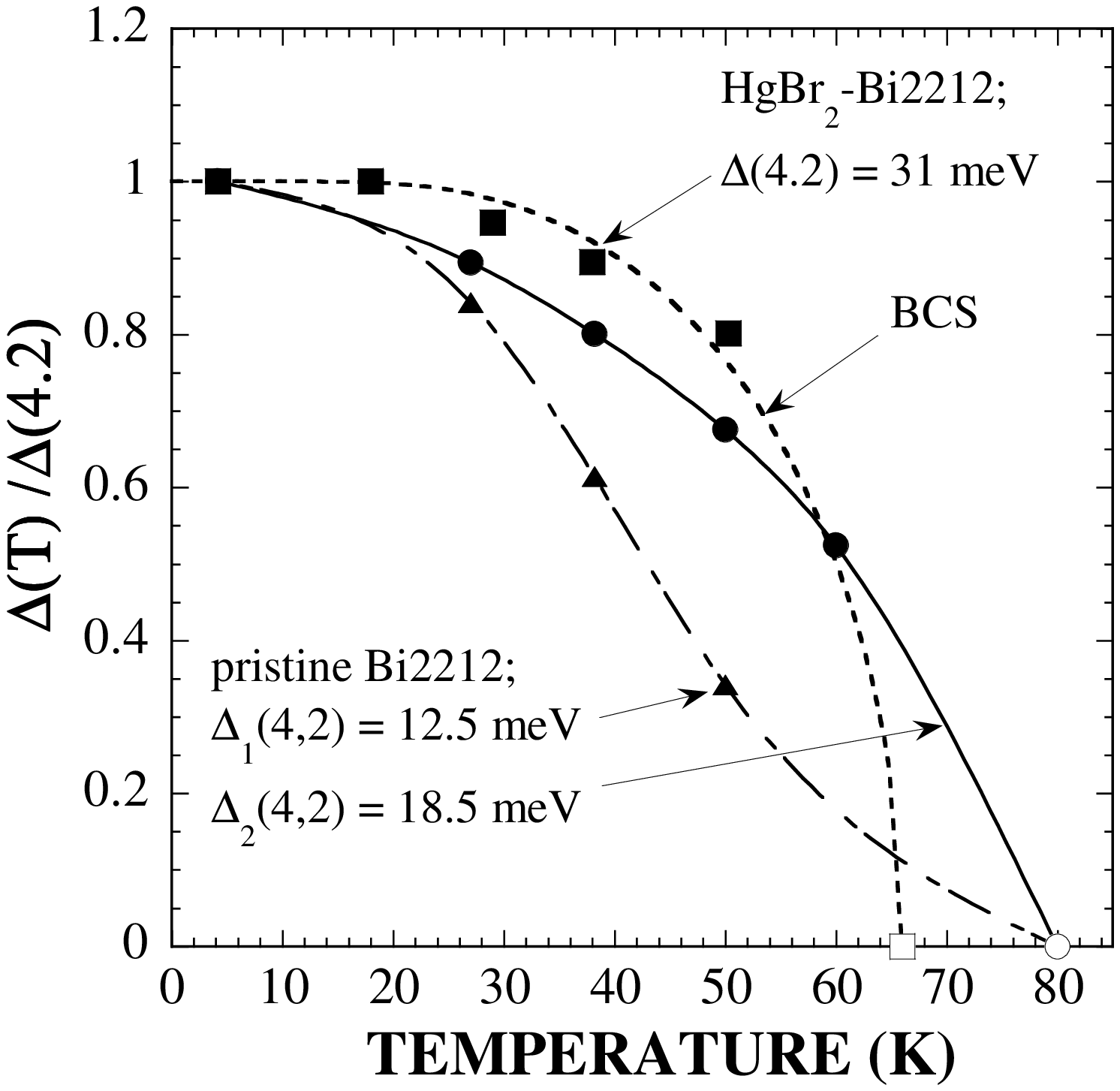,width=7.0cm}
\end{center}
\vspace{5mm}
\fcaption{The normalized energy gap parameter 
$\Delta $ vs. temperature for the pristine (circles and
triangles), and 
$\rm (HgBr_2$)Bi2212- (squares) samples. All
lines are guides for the eye.  There were two distinct gap
features for the pristine sample, $\Delta _1$ and
$\Delta _2$.  The empty symbols correspond to the
disappearance  of the {\it c}-axis critical current.
}
\label{delta}
\end{figure}

The temperature dependence of the normalized energy gap parameter
$\Delta $ is shown in Fig.~\ref{delta}.  The BCS-like temperature
dependence is shown for comparison.  For the pristine sample
there were two distinct features in d$I/$d$V(V)$-curves, which
may reflect either the in-plane anisotropy of the superconducting
energy gap parameter,\cite{Ma} or priximity effects in the complex
multilayer system of Bi2212.\cite{we_PRB} 

The superconducting energy gap cannot be traced close to $T_c$,
the corresponding feature of the d$I/$d$V(V)$-curves vanishes
earlier, although the superconducting critical current can  be
seen clearly from measured $I-V$-characteristics 
in the whole temperature range up to $T_c$, 
see closed and open symbols in Fig.~\ref{delta}. 

\section{Conclusions}
\noindent
In conclusion, we observed both superconducting and pseudogap features
of low-transparency intrinsic tunneling in Bi-2212 single crystals 
intercalated with $\rm HgBr_2$. The dramatic decrease of 
the coupling between 
CuO$_2$-bilayers in the {\it c}-axis direction after the intercalation 
does not influence much the shape of $R_c(T)$-curves. The characteristic 
temperature when the pseudogap features set in in the 
current-voltage characteristics with decreasing of temperature 
stays also nearly unchanged. All this implies that it is
properties of an individual CuO$_2$-bilayer (plus surrounding charge
reservoirs) which determine both $R_c(T)$ and the pseudogap in
the single particle spectrum of Bi2212.

\nonumsection{References}


\noindent
\begin{thebibliography}{000}
\bibitem{all}
W. W. Warren \textit{et al}.,
{\nineit Phys. Rev. Lett.} {\ninebf 62}, 1193 (1989);
G. V. M. Williams \textit{et al}.,
{\nineit Phys. Rev. Lett.} {\ninebf 78}, 721 (1997);
R. Nemetschek \textit{et al}.,
{\nineit Phys. Rev. Lett.} {\ninebf 78}, 4837 (1997);
J. W. Loram \textit{et al}.,
{\nineit Phys. Rev. Lett.} {\ninebf 71}, 1740 (1993);
H. Ding \textit{et al}.,
{\nineit Nature (London)} {\ninebf 382}, 51 (1996);
M. R. Norman \textit{et al}.,
{\nineit Nature (London)} {\ninebf 392}, 157 (1998). 
See also a review paper by T. Timusk and B. Statt, 
cond-mat/9905219.

\bibitem{overdoped}
H. J. Tao, F. Lu, and E. L. Wolf,
{\nineit Physica C} {\ninebf 282-287}, 1507 (1997).

\bibitem{STM}
Ch. Renner \textit{et al}.,
{\nineit Phys. Rev. Lett.} {\ninebf 80}, 149 (1998).

\bibitem{Hasegawa}
T. Hasegawa, H. Ikuta, and K. Kitazawa,
in {\nineit Physical Properties of High Temperature Superconductors III},
ed. by D. M. Ginsberg (World Scientific,Singapore, 1992) p. 525.

\bibitem{Muller}
R. Kleiner \textit{et al}.,
{\nineit Phys. Rev. Lett.} {\ninebf 68}, 2394 (1992);

\bibitem{sub_gap}
Ch. Helm \textit{et al}.,
{\nineit Phys. Rev. Lett.} {\ninebf 79}, 737 (1997).

\bibitem{Suzuki}
M. Suzuki, T. Watanabe, and A. Matsuda,
{\nineit Phys. Rev. Lett.} {\ninebf 82}, 5361 (1999).


\bibitem{Gough}
I. F. G. Parker \textit{et al}., in
{\nineit Conference on Superlattices II: Native and Artificial},
ed. I. Bozovic and D. Pavuna,
(SPIE, Bellingham, WA, 1998), SPIE Vol. 3480, p. 11.

\bibitem{Choy}
J.-H. Choy, S.-J. Hwang, and N.-G. Park,
{\nineit J. Am. Chem. Soc.} {\ninebf 119}, 1624 (1997).

\bibitem{we_APL}
A. Yurgens \textit{et al}.,
{\nineit Appl. Phys. Lett.} {\ninebf 70}, 1760 (1997).

\bibitem{iodine} 
X.-D. Xiang \textit{et al}.,
{\nineit Nature} {\ninebf 348}, 145 (1990).

\bibitem{we_PRL2} In experiments with hydrostatic pressure
we also observed that despite significant change of
the {\nineit c}-axis resistance with pressure, 
its temperature dependence did not change much, see
A. Yurgens \textit{et al}.,
{\nineit Phys. Rev. Lett.} {\ninebf 82}, 3148 (1999). 

\bibitem{deWilde} Y. DeWilde \textit{et al}.,
{\nineit Phys. Rev. Lett.} {\ninebf 80}, 153 (1998).

\bibitem{Ma} 
J. Ma \textit{et al}.,
{\nineit Science} {\ninebf 267}, 862 (1995).

\bibitem{we_PRB}
A. Yurgens \textit{et al}.,
{\nineit Phys. Rev. B} {\ninebf 53}, R8887 (1996);



\end{thebibliography}
\end{document}